\RequirePackage{amsmath}
\documentclass[twocolumn,epjc3]{svjour3}  
\RequirePackage[T1]{fontenc}
\RequirePackage{graphicx}
\RequirePackage{mathptmx}      % use Times fonts if available on your TeX system
\RequirePackage{flushend}
\RequirePackage[numbers,sort&compress]{natbib}
\RequirePackage[colorlinks,citecolor=blue,urlcolor=blue,linkcolor=blue,unicode]{hyperref}
\usepackage{lmodern}
\usepackage{graphics}
\usepackage{amssymb,slashed,xcolor,cancel}
\usepackage{textgreek}
\journalname{Eur. Phys. J. C}
\begin{document}\sloppy

\title{Two Gluon Emission from MHV:
Two Particle Correlations \\and the Deviation from Poisson}
%\subtitle{Do you have a subtitle?\\ If so, write it here}

%\titlerunning{Short form of title}        % if too long for running head

\author{A.\ N.\ Rasoanaivo\thanksref{e1,addr1}
        \and
        W.\ A.\ Horowitz\thanksref{e2,addr1} %etc.
}

%\thankstext{t1}{Grants or other notes
%about the article that should go on the front page should be
%placed here. General acknowledgments should be placed at the end of the article.
\thankstext{e1}{e-mail: andriniaina@aims.ac.za}
\thankstext{e2}{e-mail: wa.horowitz@uct.ac.za}

%\authorrunning{Short form of author list} % if too long for running head

\institute{Department of Physics, University of Cape Town, Private Bag X3, Rondebosch 7701, South Africa \label{addr1}
}

\date{\phantom{Received: date / Accepted: date}}
% The correct dates will be entered by the editor

\maketitle

\begin{abstract}
We derive the momentum distribution for two soft and collinear gluons emitted from an off-shell quark, modeled by a $q\gamma\rightarrow qgg$ scattering process and computed using the maximal helicity violating (MHV) technique.  We find the explicit non-Abelian corrections to the independent emission Poisson distribution in QED and demonstrate the efficiency of the MHV technique in perturbative QCD calculations.   
The gluonic two particle correlation function is highly suggestive of that measured in high-multiplicity $pp$ and $pA$ collisions, with a tight correlation in angle but broad correlation in rapidity.  Further, the autocorrelation generated by the non-Abelian correlation can provide a significant non-hydrodynamical source of $v_2$ in these high multiplicity $pp$ and $pA$ collisions. Finally, these two gluon correlations can be used to calibrate Monte Carlo shower codes for a more accurate description of QCD processes in $pp$ collisions.
%\keywords{First keyword \and Second keyword \and More}
% \PACS{PACS code1 \and PACS code2 \and more}
% \subclass{MSC code1 \and MSC code2 \and more}
\end{abstract}

\newcommand{\abraket}[1]{\left \langle #1\right \rangle}
\newcommand{\sbraket}[1]{\left [ #1\right ]}
\newcommand{\tbraket}[1]{\left\langle #1\right]}
\newcommand{\tbraketbar}[1]{\left[ #1\right\rangle}
\newcommand{\trace}[1]{\textrm{tr}\left (#1\right )} 

\newcommand{\pT}{$p_T$}
\newcommand{\pt}{\pT}
\newcommand{\highpT}{high-\pT{}}
\newcommand{\highpt}{\highpT}

\newcommand{\fig}[1]{Fig.\ \ref{#1}}
\newcommand{\eq}[1]{Eq.~\ref{#1}}
\newcommand{\redvtwo}{$\overline{\abraket{v_2}}$}
\newcommand{\redvtwoN}{$\overline{\abraket{v_2}}(N)$}

\section{Introduction}
\label{intro}
Lattice QCD shows that at a temperature around $160-170$ MeV there is a crossover phase transition from normal nuclear matter to what has been called the quark-gluon plasma (QGP) \cite{Philipsen:2012nu}.  Experiments conducted at the Relativistic Heavy Ion Collider (RHIC) and the Large Hadron Collider (LHC) aim to uncover the properties of the QGP predicted by lattice QCD and created in these massive machines.

One of the most important observables used to probe the properties of the QGP is jet quenching, the suppression of particles with very high momentum in the direction transverse to the colliding hadrons.  
The idea is that quarks or gluons (collectively known as partons) of initially high transverse momentum, high-\pT{}, produced in the initial overlap of colliding hadrons lose energy as they propagate through the expanding QGP medium \cite{Majumder:2010qh}.  (Here, \highpT{} generally means $p_T\gtrsim2$ GeV $\gg\Lambda_{QCD}\sim$ 170 MeV.) Since the production spectrum of \highpT{} partons falls steeply, an energy loss of the partons leads to an observable suppression in the number of detected particles at a specific momentum compared to reference collisions without a QGP medium.  If we assume that both the production of the initial distribution of \highpT{} partons and the final fragmentation of these partons into hadrons are unchanged from the reference collisions, then a detailed comparison of the measured suppression spectrum with energy loss models based on theoretical predictions for the medium-probe dynamics allows for a determination of the properties of the QGP.

In the weakly-coupled paradigm, the \highpt{} particle is assumed to be weakly coupled to a QGP medium that is weakly coupled to itself.  Then one may use the methods of perturbative QCD (pQCD) to compute the properties of the QGP medium and its interaction with the \highpT{} probe.  In this paradigm, the momentum of the \highpt{} parton is altered by a combination of elastic and inelastic collisions with medium quasiparticles, which are slightly thermally modified free quarks and gluons \cite{Majumder:2010qh,Djordjevic:2011dd}. At relativistic speeds, radiative processes contribute to the majority of the energy lost by the \highpT{} parton \cite{Wicks:2005gt} (note, though, that collisional energy loss is still a significant fraction of the total, even out to $p_T\gtrsim200$ GeV \cite{Horowitz:2012cf}). 

Much of the two decades of theoretical effort in pQCD-based radiative energy loss calculations focused on computing the single inclusive spectrum of gluon radiation from interactions between a \highpT{} parton and a medium quasiparticle \cite{Majumder:2010qh}.  %radiative energy loss processes 
Since estimates for the number of gluons radiated by a typical \highpT{} parton in medium at RHIC or LHC is $\sim3$ \cite{Gyulassy:2001nm}, a realistic energy loss model must take into account the very likely possibility for multiple gluon emission.  
In QED, one can rigorously show that multiple soft and collinear photon emission are independent and uncorrelated and follows from the Poisson convolution of the single inclusive photon distribution \cite{Peskin:1995ev,Weinberg:1995mt}. A common assumption in pQCD-based energy loss models is that the Poisson distribution is a good approximation for the distribution of multiple radiated gluons \cite{Gyulassy:2001nm,Majumder:2010qh,Armesto:2011ht}.  
Since including multiple gluon emission is an important effect \cite{Gyulassy:2001nm,Majumder:2010qh,Armesto:2011ht}, it is important to test this Poisson approximation.

There has been recent partial progress in computing the momentum distribution of two radiated gluons from a parton undergoing multiple scattering in a QGP medium \cite{Arnold:2016kek,Arnold:2016mth,Arnold:2016jnq}. 

Holding off on the full two gluon emission calculation in medium, we fully answer a simpler question: what is the momentum distribution of two soft and collinear gluons radiated by an off-shell quark?

The above forms a first test problem as we work to gain insight into the more difficult problem of multiple gluon emission stimulated by multiple scattering in a QGP. 
There are additional benefits to examining the two gluon emission from an off-shell quark.  First, by quantifying the correlations amongst two gluons emitted that do not have strong angular ordering, we provide a benchmark for improving vacuum Monte Carlo showering programs \cite{Lonnblad:1992tz,Gleisberg:2008ta,Sjostrand:2014zea,Bellm:2015jjp}; these programs are crucially important for understanding the QCD background in particle physics, for example in beyond Standard Model (BSM) particle searches \cite{Buckley:2011ms}, and for applications in energy loss models in heavy ion physics \cite{Zapp:2008gi,Zapp:2012ak,Renk:2008pp,Armesto:2009fj}.  

Second, with our two gluon correlation function, we may make a connection with the two particle correlations measured in hadronic collisions.  Very interesting structures in the correlations of intermediate-\pT{} $\sim$ few GeV/c particles, tightly distributed in angle $\phi$ but very broad in rapidity $\eta$, have been observed at RHIC and LHC in nucleus-nucleus, $AA$, and high multiplicity proton-nucleus, $pA$, and proton-proton, $pp$, collisions \cite{Wong:2007pz,Wenger:2008ts,Abelev:2009af,Abelev:2012ola,Aad:2012gla,Khachatryan:2010gv,CMS:2012qk,Aad:2014lta,Khachatryan:2015lva}.  A number of theoretical explanations have been posited for the physics behind these correlations \cite{Romatschke:2006bb,Shuryak:2007fu,Gavin:2008ev,Dumitru:2010iy,Ozonder:2016xqn,Dusling:2017dqg}. We will show that the two gluon correlations induced by the non-Abelian structure of QCD provide a natural source for such semi-hard particle correlations.

Finally, the azimuthal anisotropy in the momentum distribution of particles in $AA$ collisions is a fertile ground for experimental and theoretical work.  In particular, the quantitative description of the second Fourier moment of this distribution, referred to as $v_2$, by nearly inviscid relativistic hydrodynamics models \cite{Teaney:2003kp,Song:2010mg,Gale:2012rq,Bernhard:2016tnd,Alqahtani:2017tnq,Weller:2017tsr} has been interpreted as implying the QGP medium is actually strongly coupled and is best described through the novel methods of the anti-de Sitter/conformal field theory (AdS/CFT) correspondence \cite{Kovtun:2004de}.  Hydrodynamics models have also recently claimed to quantitatively describe the $v_2$ in high multiplicity $pA$ and $pp$ collisions \cite{Bozek:2010pb,Werner:2010ss,Weller:2017tsr}, systems whose size has been previously thought too small for hydrodynamics to apply \cite{Romatschke:2017vte}.  However, this nearly perfect fluid/strong coupling paradigm has been challenged by parton cascade models with pQCD-like cross sections that give qualitatively similar results \cite{Molnar:2001ux,Lin:2001zk,Bzdak:2014dia,Koop:2015wea,He:2015hfa,Lin:2015ucn}. 
We will further challenge the success of this hydrodynamics description, at least in high multiplicity $pp$ and $pA$ collisions, by indicating that the autocorrelation from multiple particle emission can lead to a large non-flow $v_2$ from the non-Abelian nature of multiple gluon emission.

In order to compute the momentum distribution for two gluon emission from an off-shell quark, we introduce the use of the maximal helicity violating (MHV) technique, which we believe provides a powerful new tool for the investigation of pQCD energy loss physics. 

This paper is divided into three main parts. In sections \ref{sec:MHV1} and \ref{sec:singleemission}, we introduce the MHV technique and then, to familiarize the reader with their use, re-derive the single inclusive gluon radiation distribution from an off-shell quark using them.  
In sections \ref{sec:twogluonamp} and \ref{sec:twogluoncrosssection}, we compute the scattering amplitude for an off-shell quark to radiate two gluons and compute the non-Abelian correction for the two gluon distribution beyond the Poisson approximation. In sections \ref{sec:twogluoncorrelation} and \ref{sec:inducedv2}, we show how the two particle correlations induced by the non-Abelian nature of QCD compares to experimental measurements, and then we compute the corresponding momentum space anisotropy of particles, $v_2$, from the azimuthal anisotropy from our correlation. We then conclude our paper in section \ref{sec:conclusions}.

\section{The maximal helicity violating technique}
\label{sec:MHV1}
The scattering amplitude is one of the most fundamental objects in field theory, and the scattering amplitude allows for a connection between theoretical developments and experimental measurements. Since Feynman, amplitudes have been derived using diagrammatic methods in which locality and unitarity are manifest, but at a cost of huge redundancies. For example, in the case of a photon that decays into a quark anti-quark pair, the number of Feynman diagrams increases faster than exponentially with the number of gluons radiated ($\sim 2^n n! $ diagrams). But, surprisingly, the final expression is remarkably simple with well chosen variables \cite{Bern:2007dw}. In this section we provide a pedagogical introduction to the maximal helicity violating (MHV) technique that highly simplifies the computation of QCD amplitudes.

\subsection{Spinor helicity formalism}
We first assume that we are in an energy regime in which all particles are approximately massless.  Then, instead of the usual external data---the momenta and polarisation vectors of the on-shell incoming and out-going particles---the scattering amplitude is expressed with a set of pairs of spinor variables $(\lambda_a^i,\widetilde{\lambda}_{\dot{a}}^i)$, to be discussed further below, and helicities $h_i$ of the external particles. Here the helicities are measured with respect to a single reference direction, which may be considered either outgoing or incoming.  Throughout this work we consider the reference direction as outgoing. 

\begin{figure}
\centering
\resizebox{0.35\textwidth}{!}{
\includegraphics{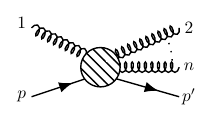} }
\caption{QCD amplitude $\mathcal{M}(p,1,2,\ldots,n,p')$ with all outgoing momenta: an anti-fermion of momentum $p$, a fermion of momentum $p'$, and $n$ gluons of momenta $k^1\ldots k^n$.}
\label{namplitude}
\end{figure}

We will find it best to classify amplitudes according to the set of helicities for all the particles in a diagram.  
A generic amplitude with an outgoing fermion, anti-fermion, and $n$ gluons is shown in \fig{namplitude}. By the helicity conservation of massless fermions, a non-zero amplitude requires the fermion and anti-fermion to have opposite helicities. Further, if all the gluons have the same helicity, then, again, the amplitude must vanish \cite{ZeeQFT2010,Henn:2014yza}. 
The first non-vanishing contribution to the amplitude comes from the so-called maximal helicity violating (MHV) diagram, in which all gluons but one have positive helicity\footnote{In our convention of taking the reference momentum for determining the helicity as outgoing, the non-vanishing amplitude in which all gluons but one have negative helicity is known as $\overline{\mathrm{MHV}}$; the $\overline{\mathrm{MHV}}$ amplitude evaluates to the same as the MHV amplitude, but with square brackets replacing angle brackets.}.  The next to maximal helicity violating (NMHV) diagram has all gluons but two with the same helicity.  And so on for NNMHV, etc.\ \cite{britto2011introduction}.
 
As mentioned above, amplitudes organized by helicity can be expressed in a very simple form, especially for MHV amplitudes: for a pure gauge amplitude the MHV amplitude is given by the Parke-Taylor formula \cite{Parke:1986gb}. The crucial ingredient for this simplification is the spinor helicity formalism. In the spinor helicity formalism, the four momentum $p_\mu$ is represented as a two by two matrix  that carries spinor indices ($P_{a\dot{a}}$); i.e.\ we take advantage of the so-called spinor map between SL$(2,\mathbb{C})$ and the proper, orthochronous Lorentz group SO$^+(3,1)$. Then the invariant mass squared $p^2=m^2$ is given by $\det (P_{a\dot{a}})$, the determinant of the matrix. 

In the case of massless particles we have $ \det(P_{a\dot{a}})=0$; since any two by two matrix with a vanishing determinant can be written as a product of two vectors, then the momentum can be factorized as
\begin{equation}
 P_{a\dot{a}}=(\sigma_{a\dot{a}})^\mu p_\mu \xrightarrow{\quad \det(P_{a\dot{a}})= 0\quad} P_{a\dot{a}}=\lambda_a\widetilde{\lambda}_{\dot{a}},
 \end{equation} 
where $\sigma^\mu=(1,\vec{\sigma})$ are the Pauli matrices, and  $\lambda_a$ and $\widetilde{\lambda}_{\dot{a}}$ are two vectors that transform respectively as left handed and right handed spinors under Lorentz transformations. 

The new variables are known as two component spinor variables $(\lambda_a,\widetilde{\lambda}_{\dot{a}})$, and we can make two different Lorentz invariant products out of them.  For two four momenta $k^1$ and $k^2$ we have
\begin{equation}
\left \{\begin{aligned}
&\abraket{12}\equiv \abraket{k^1k^2}=\epsilon^{ab}\lambda_a^1\lambda^2_b\\&\sbraket{12}\equiv \sbraket{k^1k^2}=\epsilon^{\dot{a}\dot{b}}\widetilde{\lambda}_{\dot{a}}^1\widetilde{\lambda}_{\dot{b}}^2,
\end{aligned}\right .
\end{equation}
where $\epsilon^{ab}$ and $\epsilon^{\dot{a}\dot{b}}$ are the Levi-Civita symbols with $\epsilon^{12}=\epsilon^{\dot{1}\dot{2}}=1$; the angle and square brackets are thus antisymmetric. 

In terms of the spinor helicity variables, MHV amplitudes are expressed only in terms of angle brackets. 
The amplitude in \fig{namplitude} is then
\begin{equation}
\mathcal{M}_{\textrm{MHV}}=\tilde{g}^n\sum_{\{1,\ldots,n\}}\frac{(T_{a_1}\cdots T_{a_n})\abraket{pI}^3\abraket{p'I}}{\abraket{p'p}\abraket{p1}\abraket{12}\cdots\abraket{n -1,n}\abraket{n p'}},
\label{MHVformula}
\end{equation}
as in \cite{Elvang:2015rqa,Maltoni:2002mq}, where the sum over $``\{1,\ldots,n\}"$ is the sum over all permutations of the gauge field indices,  $I$ is the index associated with the one gluon of different helicity, $T_{a_i}$ is a generator of $SU(N)$ and is associated with the $i$-th particle, and $\tilde{g}=g\sqrt{2}$ is the QCD coupling constant. For the full cross section, we will square the amplitude with its complex conjugate and average over initial states and sum over final states as per usual over all helicity configurations. At the level of MHV diagrams, this summing and averaging corresponds with summing over the $I$ index.  Amplitudes with less helicity violation, NMHV, NNMHV, etc., can be generated from the  Britto-Cachazo-Feng-Witten (BCFW) recursion relation \cite{Britto:2005fq}. However, we will show below that for the soft-collinear radiation we're interested in, we need only consider MHV diagrams. 
\subsection{Useful relations}
\label{useful}
We expect observable results to depend only on scalar products of external momenta.  As we square our amplitudes and sum over initial and final states, we will find the following relations useful:
\begin{itemize}
\item Complex conjugate: $\abraket{12}^*=\sbraket{12}$.

\item Scalar product: $\abraket{12}[12]=2k_1. k_2$.

\item Schouten identity \cite{ZeeQFT2010,Henn:2014yza}: $$\abraket{12}\abraket{34}+\abraket{13}\abraket{42}+\abraket{14}\abraket{23}=0.$$

\item Relation to Feynman slash \cite{Henn:2014yza}: $$\abraket{12}\sbraket{23}\equiv\tbraket{123}=\langle1|\slashed{k}_2|3].$$

\item Trace of four gammas, as shown in the appendix:$$\tbraket{12341}+\tbraket{12341}^*=\trace{\slashed{k}_1\slashed{k}_2\slashed{k}_3\slashed{k}_4}.$$
\end{itemize}

With the MHV technique now introduced, we see the simplicity of using the spinor variables: we have a formula, \eq{MHVformula}, that allows us to compute $n$ gluon emission amplitudes which are almost impossible to compute using the usual diagrammatic techniques. In the next section we see how we may derive expressions for soft-collinear radiation using MHV. 

\section{Single gluon production}
\label{sec:singleemission}
\begin{figure}[!htbp]
\centering
\resizebox{0.4\textwidth}{!}{
\includegraphics{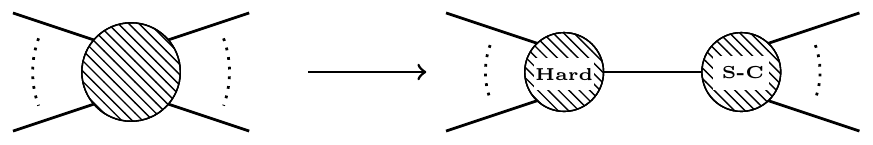}}
\caption{Visualization of the soft-collinear factorization of amplitudes due to the Weinberg Soft Theorem \protect\cite{Bianchi:2014gla}.}
\label{Diagram_factorization}
\end{figure}

As a pedagogical example for using the spinor helicity formalism, in this section we reproduce the well-known result for single gluon emission using MHV. 
We start with the Weinberg Soft Theorem \cite{Bianchi:2014gla}, which states that for a given process the soft-collinear radiation does not interfere with the hard scattering that produces it because the two processes occur at different energy scales.  Thus the diagram for a generic process that has both a hard scattering and soft-collinear emission factorizes as shown in \fig{Diagram_factorization}. 

\begin{figure}[!htbp]
\centering
\resizebox{0.4\textwidth}{!}{
\includegraphics{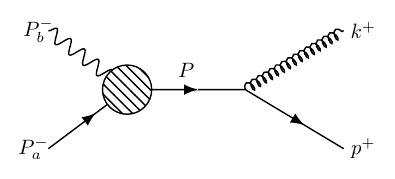}}
\caption{Single gluon emission from an on-shell quark struck by an on-shell photon. Note that all helicities are measured against a reference direction that is considered outgoing.}
\label{Single_emission}
\end{figure}

In this work, we are interested in a highly energetic parton with small virtuality or, equivalently, a highly energetic parton emerging from a hard process; thus the parton will be able to radiate gluons. For our purposes, let the parton be a quark produced by a hard photon-quark interaction; see \fig{Single_emission}.  Our imagined process is a simplified version of hard parton production in hadronic collisions or is a model for nearly on-shell photon interaction with hadrons in an electron-ion collider (EIC).  

Now, having that picture in mind, we can set our variables. For the hard scattering, let $P_a^\mu$ be the momentum of the incoming quark with helicity $-\frac{1}{2}$ and $P_b^\mu$ for the photon with helicity $-1$. Let $P^\mu$ be the momentum of the quark after the photon is absorbed, and this quark will radiate gluons in the soft-collinear region. For the soft-collinear process $k^\mu$ will be the momentum of the gluon of helicity $h$ and $p^\mu$ the momentum of the quark final state of helicity $+\frac{1}{2}$. Using the MHV amplitude \eq{MHVformula}, we can derive 
\begin{align}
\mathcal{M}_{\textrm{\tiny MHV}}&=e \tilde{g}\sum_{\textrm{perm}}\frac{T_{a_b}T_{a_g}\abraket{ab}^3\abraket{pb}}{\abraket{a b}\abraket{b k}\abraket{k p} \abraket{p a}}\nonumber\\
&=e \tilde{g}\frac{\abraket{ab}^3\abraket{pb}}{\abraket{p a}}\left (\frac{T_{a_g}}{\abraket{a b}\abraket{b k}\abraket{k p} }+\frac{T_{a_g}}{\abraket{a k}\abraket{k b}\abraket{b p} }\right )\nonumber\\
&= \left (e\frac{\abraket{ab}^3\abraket{pb}}{\abraket{a b}\abraket{bp}\abraket{pa}}\right )\left (\tilde{g} T_{a_g}\frac{\abraket{ap}}{\abraket{a k}\abraket{k p} }\right )
\label{derivationOfSingle}
\end{align}
In the above, the first line is a direct application of the MHV amplitude in \eq{MHVformula}, where we sum over the permutation of both the gauge fields, the photon and the gluon.  $T_{a_b}$ is the color associated with the photon, which is set to one, $T_{a_b}=1$, since photons are color blind and are not in the representation of $SU(N)$. In the second line we write explicitly the sum over the permutation; to get to the third line, we decouple the kinematics of the photon and the gluon by putting them in a common denominator then simplifying using the Schouten identity. In the third line we group the expression into two parts: in the first part we have the hard process that contains the kinematics of the photon; in the second part we have the color and kinematics of the soft-collinear gluon, which we denote by the eikonal factor,
\begin{equation}
  J_1(k) \equiv \tilde{g} T_{a_g}\frac{\abraket{ap}}{\abraket{a k}\abraket{k p} }.
 \end{equation} 
Notice how the single inclusive amplitude above, \eq{derivationOfSingle}, naturally factorizes as it must according to the Weinberg Soft Theorem \cite{Bianchi:2014gla}.

We may also express the eikonal factor $J_1(k)$ above in terms of $P_\mu$ using momentum conservation $P_a^\mu=P^\mu-P_b^\mu$
\begin{equation}\begin{aligned}
\frac{\abraket{ap}}{\abraket{ak}}& =\frac{\sbraket{ba}}{\sbraket{ba}}\frac{\abraket{ap}}{\abraket{ak}}=\frac{\tbraketbar{bap}}{\tbraketbar{bak}}=\frac{[b|\slashed{a}|p\rangle}{[b|\slashed{a}|k\rangle}\\&=\frac{[b|(\slashed{P}-\slashed{b})|p\rangle}{[b|(\slashed{P}-\slashed{b})|k\rangle}=\frac{\sbraket{bP}\abraket{Pp}-\sbraket{bb}\abraket{bp}}{\sbraket{bP}\abraket{Pk}-\sbraket{bb}\abraket{bk}}\\& =\frac{\abraket{Pp}}{\abraket{Pk}}.
\end{aligned}
\end{equation}
Then the eikonal factor can be expressed in terms of $P^\mu$ instead of $P_a^\mu$, and the color average of $J_1$ squared is given by
\begin{equation}
\begin{aligned}
\label{single}
\overline{\left |J_1(k)\right |^2}&=\trace{\left |\tilde{g} T_{a_g}\frac{\abraket{Pp}}{\abraket{Pk}\abraket{kp}}\right |^2}\\
&=8\pi \alpha_sC_AC_F \frac{\abraket{Pp}\sbraket{Pp}}{\abraket{Pk}\sbraket{Pk}\abraket{kp}\sbraket{kp}}\\
&=4\pi \alpha_sC_AC_F \frac{P. p}{(P. k)(p. k)},
\end{aligned}
\end{equation}
where $\alpha_s=g^2/(4\pi)$, $C_A=N$, and $C_F=(N^2-1)/(2N)$. \eq{single} contains the usual single inclusive emission distribution kinematics given a hard scattering, multiplied by the appropriate gauge invariants \cite{Peskin:1995ev}.

For the probability $d\mathcal{W}$ of emitting a single gluon with a momentum $k$ given the hard photon scattering process (i.e.\ with the hard photon kernel divided out), we must average over the initial states and sum over the final states which will give us an extra factor of 2.  Since the amplitude \eq{derivationOfSingle} is independent of the initial helicity, and also fixed once the initial helicity is given, we must only divide out by the initial number of possible colors for the incoming quark; i.e.\ we must divide by $C_A$.  Then we have

\begin{equation}
\label{onegluon}
d\mathcal{W}(k)=\frac{d^3k}{(2\pi)^3}\frac{1}{2 E_k} 8\pi \alpha_s \frac{C_F P. p}{(P. k)(p. k)}.
  \end{equation} 
  
\section{Amplitude for \texorpdfstring{$q\gamma\to qgg$}{q\textgamma -> qgg}}
\label{sec:twogluonamp}
\begin{figure}[!htbp]
\centering
\resizebox{0.3\textwidth}{!}{
\includegraphics{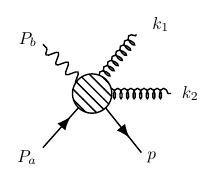} }
\caption{Generic Feynman diagram for $q\gamma\to qgg$.}
\label{5amplitude}
\end{figure}

Having demonstrated the utility of MHV for the case of single gluon emission, let us now apply MHV to two gluon emission; see \fig{5amplitude}.  We will continue to find it useful to categorize amplitudes according to their helicities.  In particular, if we denote a particular amplitude as $\mathcal{M}_{h_\gamma h_{1} h_{2}}$ 
with $h_\gamma$ the helicity of the photon and  $h_1$ and $h_2$ the helicities of the radiated gluons, then we have that
\begin{align}
  \overline{|\mathcal{M}|^2}
  &=\sum_\textrm{helicities}|\mathcal{M}(\textrm{helicities})|^2\nonumber\\
  &=2(|\mathcal{M}_{-++}|^2+|\mathcal{M}_{+-+}|^2+|\mathcal{M}_{++-}|^2)\\
  &=\sum_{\{\textrm{MHV}\}} |\mathcal{M}_\textrm{MHV}|^2.\nonumber
\end{align}
Note that in going from the first line to the second, we have used the facts that 1) $\mathcal{M}(+++)=\mathcal{M}(---)=0$ and 2) $\mathcal{M}(\textrm{all helicities flipped})=\mathcal{M}^*(\textrm{helicities})$; i.e.\  in a given amplitude
all the helicities can be flipped by complex conjugation \cite{Ozeren:2006ft}.  For the last line, we recall that an MHV amplitude is one in which all but one helicity is the same.

In order to derive an expression for the cross section for two gluon emission from a hard quark scattering, we therefore need to compute the various MHV configurations for $q\gamma\rightarrow qgg$.  Starting from \eq{MHVformula}, we have 
\begin{align}
\label{double}
\mathcal{M}_\textrm{MHV}&=e\tilde{g}^2\frac{\abraket{aI}^3\abraket{pI}}{\abraket{ap}}\sum_{\{b,1,2\}}\frac{T_{a_b}T_{a_1}T_{a_2}}{\abraket{ab}\abraket{b1}\abraket{12}\abraket{2p}}\nonumber\\
%&=e\tilde{g}^2\frac{\abraket{aI}^3\abraket{pI}}{\abraket{pa}}\sum_{\{1,2\}}\left (\frac{T_{a_b}T_{a_1}T_{a_2}}{\abraket{ab}\abraket{b1}\abraket{12}\abraket{2p}}\right .\\&\left .\hspace{0.5cm}+\frac{T_{a_1}T_{a_b}T_{a_2}}{\abraket{a1}\abraket{1b}\abraket{b2}\abraket{2p}}+\frac{T_{a_1}T_{a_2}T_{a_b}}{\abraket{a1}\abraket{12}\abraket{2b}\abraket{bp}}\right )\\
&=e\tilde{g}^2\frac{\abraket{aI}^3\abraket{pI}}{\abraket{ap}}\sum_{\{1,2\}}\left (\frac{T_{a_1}T_{a_2}}{\abraket{ab}\abraket{b1}\abraket{12}\abraket{2p}}\right .\nonumber\\&\left .\hspace{0.5cm}+\frac{T_{a_1}T_{a_2}}{\abraket{a1}\abraket{1b}\abraket{b2}\abraket{2p}}+\frac{T_{a_1}T_{a_2}}{\abraket{a1}\abraket{12}\abraket{2b}\abraket{bp}}\right )\nonumber\\
%&=e\tilde{g}^2\frac{\abraket{aI}^3\abraket{pI}}{\abraket{pa}}\sum_{\{1,2\}}\frac{\abraket{ap}}{\abraket{ab}\abraket{bp}}\frac{T_{a_1}T_{a_2}}{\abraket{a1}\abraket{12}\abraket{2p}}\\
&=e\tilde{g}^2\frac{\abraket{aI}^3\abraket{pI}}{\abraket{ab}\abraket{bp}\abraket{pa}}\sum_{\{1,2\}}\frac{T_{a_1}T_{a_2}\abraket{ap}}{\abraket{a1}\abraket{12}\abraket{2p}}.
\end{align}
\eq{double} is derived in a way very similar to the single gluon expression, \eq{derivationOfSingle}. The first line is a direct application of the fundamental formula, \eq{MHVformula}, where $I$ stands for the momentum of the boson with negative helicity.  In the second line, we first set $T_{a_b}=1$ for the photon.  Then we wrote out the various permutations of the photon, thus reducing the sum over permutations of the photon and two gluons to just a sum over the permutations of the two gluons.  Finally, for the third line, we added the three terms in the parentheses by finding a common denominator, applied the Schouten identity twice, and canceled common factors. 

Since our main interest here is the soft collinear emission of two gluons with respect to the outgoing quark, where $\abraket{p1}$ and $\abraket{p2}$ go to zero, then let us introduce a dimensionless quantity that can measure the strength of the individual MHV amplitude. Let us call this quantity the ``MHV charge,'' $Q_\textrm{MHV}$, defined by 
\begin{multline}
%\mathcal{M}_\textrm{MHV}\equiv Q_\textrm{MHV} e\frac{\abraket{ab}^3\abraket{pb}}{\abraket{ab}\abraket{bp}\abraket{pa}} \sum_{\{1,2\}}\frac{\tilde{g}^2T_{a_1}T_{a_2}\abraket{ap}}{\abraket{a1}\abraket{12}\abraket{2p}},
\mathcal{M}_\textrm{MHV}\equiv Q_\textrm{MHV}\times e\frac{\abraket{ab}^3\abraket{pb}}{\abraket{ab}\abraket{bp}\abraket{pa}}\times \\ \sum_{\{1,2\}}\frac{\tilde{g}^2T_{a_1}T_{a_2}\abraket{ap}}{\abraket{a1}\abraket{12}\abraket{2p}},
% \mathcal{M}_\textrm{MHV}\equiv Q_\textrm{MHV}\left(e\frac{\abraket{ab}^3\abraket{pb}}{\abraket{ab}\abraket{bp}\abraket{pa}}\right) \sum_{\{1,2\}}\frac{\tilde{g}^2T_{a_1}T_{a_2}\abraket{ap}}{\abraket{a1}\abraket{12}\abraket{2p}},
\label{2emissionFactorization}
\end{multline}
where we have introduced the hard scattering Born amplitude into our amplitude.  For momentum $I$ with negative helicity, we thus have
\begin{equation}
Q_\textrm{MHV}=\frac{\abraket{aI}^3\abraket{pI}}{\abraket{ab}^3\abraket{pb}}.
\end{equation}

From the definition of the MHV charge, we can see that the charge associated with the case in which the photon has a negative helicity is  $Q(-++)=1$. On the other hand, in the collinear region $p. k_1$ and $p. k_2$ are small compared to $P_a. P_b$, etc. Thus the two other cases $Q(+-+)$ and $Q(++-)$ are small since they are proportional to  $\abraket{p1}$ and $\abraket{p2}$, respectively: 
\begin{equation}
\left \{\begin{aligned}
 Q(-++)&=1,\\
 Q(+-+)&\ll1,\\
 Q(++-)&\ll1.
\end{aligned}\right .
\end{equation}  

\begin{figure}[!htbp]
\centering
\resizebox{0.4\textwidth}{!}{
\includegraphics{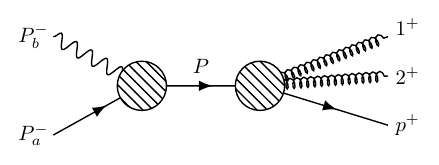}}
\caption{Gluon radiation from an electromagnetic interaction.
\label{GluonRadiation}}
\end{figure}

Therefore the only amplitude that will contribute to two gluon emission at leading order is the MHV amplitude in which the photon has one helicity and the two emitted gluons have helicities opposite that of the photon: all other amplitudes are suppressed in the soft-collinear region of phase space we are interested in; see \fig{GluonRadiation}.  Notice that our result for two gluon emission also satisfies the Weinberg Soft Theorem \cite{Bianchi:2014gla}. 

%%%%%%%%%%%%%%%%%%%%%%%%%%%%%%%%%%%%%%%%%%%%%%%%%%%%%%%%%%%%%%%%%%%%%%%%%%%%%%%%%%%%%%%%%%%%%%%%%%%%%%%%%%%%%%%%%%%%%%%%%%%%%%%%%%%%%%%%%%%%%%%%
\section{Cross section for two gluon emission}
\label{sec:twogluoncrosssection}  
In order to compute the cross section for two gluon emission, we must compute the square of the eikonal factor for two gluon emission, $J_2$.  

Since for soft-collinear emission, the dominant contribution comes from the configuration with both gluons having the same helicity, which is opposite that of the photon, we may find $J_2$ directly from \eq{2emissionFactorization} with MHV charge $Q_\textrm{MHV}=1$: 
\begin{equation}
J_2(1,2)=\tilde{g}^2\left (\frac{T_{a_1}T_{a_2}\abraket{Pp}}{\abraket{P1}\abraket{12}\abraket{2p}}+\frac{T_{a_2}T_{a_1}\abraket{Pp}}{\abraket{P2}\abraket{21}\abraket{1p}}\right ).
\label{two_gluon_emissions}
\end{equation}

Squaring $J_2$ directly makes the computation complicated since the color factors and the kinematics will mix. To avoid such mixing, we rewrite \eq{two_gluon_emissions} as a sum over the different symmetries under gluon permutation. In the case of two gluon emission, there are two possible configurations: symmetric and antisymmetric under interchange of gluons. 
We may thus write
\begin{equation}
J_2\equiv\frac{\tilde{g}^2}{2}C_{1,2}^s J^s_2(1,2)+\frac{\tilde{g}^2}{2}C_{1,2}^aJ^a_2(1,2),
\label{SymmetryExpansion}
\end{equation}
where $C^s_{1,2}$ and $C^a_{1,2}$ are the symmetrized and antisymmetrized color factors and $J_2^s(1,2)$ and $J_2^a(1,2)$ are the symmetrized and antisymmetrized kinematics of the eikonal factor:
\begin{itemize}
\item[$\bullet$] Color factors
\begin{equation}
\left\{\begin{aligned}
&C^s_{1,2}=T_{a_1}T_{a_2}+T_{a_2}T_{a_1}=\{T_{a_1},T_{a_2}\}\\[0.1cm]
&C^a_{1,2}=T_{a_1}T_{a_2}-T_{a_2}T_{a_1}=[T_{a_1},T_{a_2}].
\end{aligned}\right.
\end{equation}
\item[$\bullet$] Kinematics
\begin{equation}
\left\{
\begin{aligned}
&J^s_2(1,2)=\frac{\abraket{Pp}}{\abraket{P1}\abraket{12}\abraket{2p}}+\frac{\abraket{Pp}}{\abraket{P2}\abraket{21}\abraket{1p}}\\%=\prod_{I=\{1,2\}}\frac{\abraket{Pp}}{\abraket{P1}\abraket{1p}}
&J^a_2(1,2)=\frac{\abraket{Pp}}{\abraket{P1}\abraket{12}\abraket{2p}}-\frac{\abraket{Pp}}{\abraket{P2}\abraket{21}\abraket{1p}}.
\end{aligned}\right.
\end{equation}
\end{itemize}
Now the trace of the square of $J_2$ is just sum of the trace squared of the individual term ($s$ and $a$); i.e.\ the cross terms vanish. Then the average square of $J_2(1,2)$ will be the square of the individual terms ($C^{s/a}$ and $J^{s/a}$).

\subsection{Symmetric configuration}

The average over initial colors and sum over final colors in the symmetric configuration yields
\begin{equation}
\trace{|C^s_{1,2}|^2}=\trace{\{T_{a_1},T_{a_2}\}^2}=4C_AC_F^2-C_A^2C_F.
\end{equation}
The square of the kinematic part is just the product of single gluon emissions, 
\begin{equation}
\left |J^s_2(1,2)\right |^2=\prod_{i=\{1,2\}}\frac{P. p}{2(P. k_i)(p. k_i)}.
\label{QEDlike}
\end{equation}
The symmetric result can be understand from the so called photon decoupling; see \cite{Ozonder:2016xqn}.  Notice that symmetrizing the kinematics is the same as replacing all the generators of $SU(N)$ in \eq{two_gluon_emissions} with identity matrices which leads to the QED like bihavior in \eqref{QEDlike}.

\subsection{Antisymmetric configuration}
The average over initial colors and sum over final colors in the antisymmetric configuration yields
\begin{equation}
\trace{|C^a_{1,2}|^2}=-\trace{[T_{a_1},T_{a_2}]^2}=C_A^2C_F.
\end{equation}
Since the symmetric part only contains an Abelian contribution, we expect information on the non-Abelian QCD nature of gluon coherence effects to emerge from the antisymmetric kinematics.  Squaring the antisymmetric kinematic piece we find
\begin{equation}
\begin{aligned}
\left |J^a_2(1,2)\right |^2
&=\left |\frac{\abraket{Pp}}{\abraket{P1}\abraket{12}\abraket{2p}}-\frac{\abraket{Pp}}{\abraket{P2}\abraket{21}\abraket{1p}}\right |^2\\
&=\frac{P. p(P.k_1 p. k_2+P. k_2 p. k_1)}{4(k_1. k_2)(P. k_1)(P. k_2)(p. k_1)(p. k_2)} \\
&+\frac{P. p\big( \langle P1p2P| + [P1p2P\rangle \big)}{8(k_1. k_2)(P. k_1)(P. k_2)(p. k_1)(p. k_2)} \\
&=\left (1+\frac{\trace{\slashed{P}\slashed{k}_1\slashed{p}\slashed{k}_2}}{2k_1.k_2P.p}\right )\!\!\prod_{i=\{1,2\}}\!\!\frac{P. p}{2P.k_ip.k_i},
\end{aligned}
\label{KinematicAntisymmetric}
\end{equation}
where we factor out a common denominator in the second line and, in the third line, we apply our ``complex conjugate'' and ``trace of gammas'' results from Section \ref{useful} and rearrange the remaining terms. 

\subsection{Cross section}
With the above symmetric and antisymmetric results in hand, we may combine everything into a total cross section for the emission of two gluons,
\begin{equation}
d^2\mathcal{W}(1,2)=\left [1+\mathcal{I}(1,2)\right ]d\mathcal{W}(1)d\mathcal{W}(2),
\label{correcteddistribution}
\end{equation}
where
\begin{equation}
\begin{aligned}
\mathcal{I}(1,2)&\equiv\left(\frac{1}{2}\frac{C_A}{C_F}\right)\frac{\trace{\slashed{P}\slashed{k}_1\slashed{p}\slashed{k}_2}}{4(k_1. k_2)(P. p)}.
\end{aligned}
\label{GluonCorrelation}
\end{equation}
Since $d\mathcal{W}(i)$ is the one gluon emission probability given by \eq{onegluon}, $\mathcal{I}(1,2)$ measures the correlation between the two emitted gluons.  $\mathcal{I}(1,2)$ is then main result of this paper: it gives the deviation away from the usual Poisson independent emission assumption due to non-Abelian QCD effects.

\section{Two gluon correlations}
\label{sec:twogluoncorrelation}
In this section we would like to understand more fully our main result, $\mathcal{I}(1,2)$, and the phenomenological consequences for the correlations in the emission of two gluons from an off-shell quark.
In statistics, the correlation function between two emission events is defined by the ratio between the probability of two emissions and the product of the probability of two independent single emissions,
\begin{align}
  \frac{d^2\mathcal{W}(1,2)}{d\mathcal{W}(1)d\mathcal{W}(2)}-1 = \mathcal{I}(1,2).
\end{align}
We see that the $\mathcal{I}(1,2)$ defined in the previous section is precisely the two gluon emission correlation function.  Note that in the definition of the correlation function we subtracted the uncorrelated part in order to make the correlation vanish in the Abelian limit. 

\subsection{Color dependence of the two gluon correlation function}
\begin{figure}[!htbp]
\centering
\resizebox{0.35\textwidth}{!}{
\includegraphics{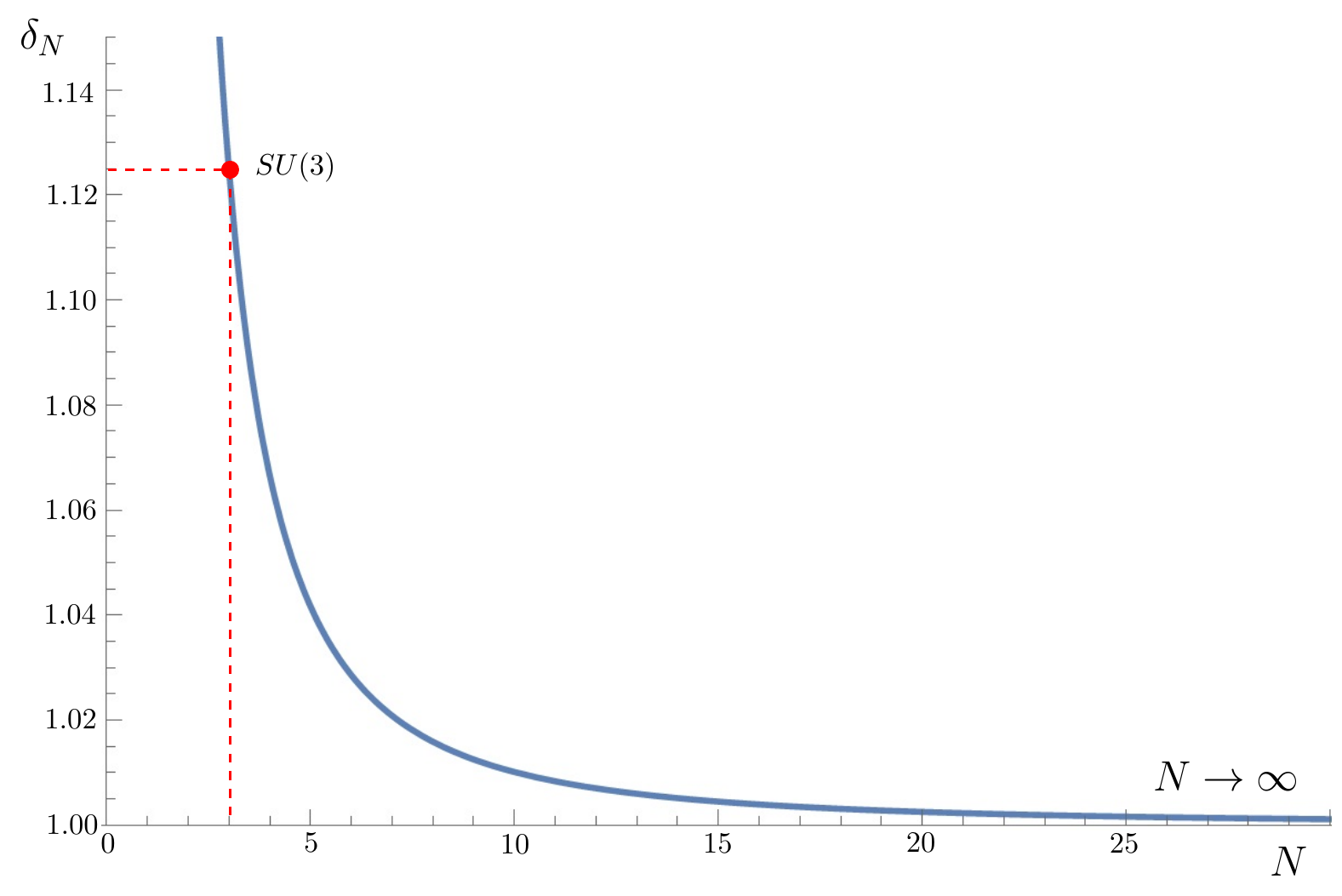}}
\caption{Correlation color factor $\delta_N$, \protect\eq{eq:deltaN}, for $N\ge 2$.}
\label{colorfactorcorrelation}
\end{figure}

We absorb the color dependence of the correlation function $\mathcal{I}(1,2)$ into a factor $\delta_N$ that we define as
\begin{equation}
\label{eq:deltaN}
\delta_N\equiv\frac{1}{2}\frac{C_A}{C_F}=\frac{N^2}{N^2-1}.
\end{equation}
In the Abelian limit $\delta_N$ vanishes since $C_A\to 0$.  As one can see in \fig{colorfactorcorrelation}, $\delta_N\rightarrow1$ as $N$ goes to infinity. For $N=3$, specific for QCD, the color factor is equal to $\delta_3=1.125$. This is to say that from $N=3$ the correlation function doesn't change much as $N$ grows; for the rest of this work we will take $\delta_N\approx 1$.

\subsection{Kinematics of the two gluon correlation function}
We can see that, as expected, $\mathcal{I}(1,2)$ is invariant under gluon exchange ($1\leftrightarrow 2$) and is singular as the two gluons become collinear.  We would like to now show that the correlation function is conformally invariant, i.e.\ invariant under the rescaling of momenta.  In order to show this symmetry, let us first evaluate the trace in terms of scalar products:
\begin{equation}
\begin{aligned}
\mathcal{I}(1,2)&= \frac{\trace{\slashed{P}\slashed{k}_1\slashed{p}\slashed{k}_2}}{4(k_1. k_2) (P. p)}\\&=\frac{(P. k_1)(p. k_2)+(P. k_2)(p. k_1)}{(k_1. k_2)(P. p)}-1.
\end{aligned}
\label{GluonCorrelation2}
\end{equation}

\begin{figure}[!htbp]
\centering
\resizebox{0.4\textwidth}{!}{
\includegraphics{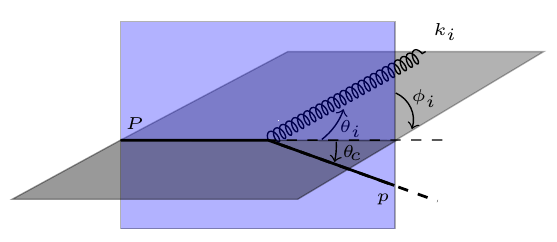}}
\caption{Angle parameters for evaluating the two particle correlation $\mathcal{I}(1,2)$, Eq.\ (\protect\ref{GluonCorrelation2}). $\phi_i$ is the angle the (blue) plane in which the gluon is emitted makes with respect to the (gray) plane in which the off-shell quark of momentum $P$ is scattered.  $\theta_i$ is the angle the emitted gluon makes with respect to the $z$ axis defined by the direction of motion of the off-shell quark of momentum $P$; similarly, $\theta_c$ is the angle the scattered quark makes with respect to the $z$ axis.}
\label{angles}
\end{figure}

In order to evaluate these dot products, let us define our variables.  
We define the momentum of the off-shell parent quark as $P^\mu\equiv(\omega_P,\vec{P})^\mu$.  After the emission of the soft-collinear photon, the now on-shell quark momentum is $p^\mu=(\omega_p,\vec{p})^\mu$.  The deflected quark makes an angle $\theta_c$ with respect to the $z$ axis defined by the direction of motion of the off-shell quark of momentum $P$; $\vec{P}.\vec{p}=|\vec{P}||\vec{p}|\cos\theta_c$.  On the other hand, the $i^\mathrm{th}$ emitted gluon has a momentum $k^\mu_i=(\omega_i,\vec{k}_i)^\mu$. The emitted gluon makes an angle $\theta_i$ with respect to the $z$ axis defined by the off-shell incoming quark and is emitted in a plane at an angle $\phi_i$ with respect to the plane defined by the incoming off-shell quark and the final on-shell quark.  See \fig{angles}.

In order to parameterize the momenta of the soft-collinear gluons in terms of the angles defined above, let us define the $z$ axis as the spatial direction of motion of the incoming, off-shell quark and the $x-z$ plane as the plane spanned by the spatial momenta of the incoming quark and the outgoing quark.  Then the momenta of the emitted gluons can be written as
\begin{equation}
\vec{k}_i=\omega_i(\sin\theta_i\cos\phi_i\,\hat{x}+\sin\theta_i\sin\phi_i\,\hat{y}+\cos\theta_i \,\hat{z}).
\end{equation}
In terms of these angles, one finds that
\begin{equation}
\begin{aligned}
\frac{(P. k_1)(p. k_2)}{(k_1. k_2)(P. p)}=&\frac{1-\cos\theta_c\cos\theta_2-\sin\theta_c\sin\theta_2\cos\phi_2}{1-\cos\theta_1\cos\theta_2-\sin\theta_1\sin\theta_2\cos\Delta\phi}\\&\times\left (\frac{1-\cos \theta_1}{1-\cos \theta_c} \right ),
  \end{aligned}
  \label{deltaphi}
\end{equation}
where $\Delta\phi$ is the azimuthal angle difference between the two gluons. We know that the direction of parent quark does not change much after radiating away soft-collinear gluons, which is the case we consider here. However, we can see from \eq{GluonCorrelation2} that the correlation function is singular if the parton is not deflected, that is to say $\theta_c$ goes to zero. We can also see that the correlation function is singular when the two gluons are exactly collinear.  We can make further progress if we isolate the $\theta_c$ and gluon collinear divergences into separate functions, which we will denote $f(1,2,\theta_c)$ and $g(1,2)$, as follows: 
\begin{equation}
\left \{\begin{aligned}
&f(1,2,\theta_c)\equiv\frac{1-\cos \theta_1}{1-\cos \theta_c}\Big(1-\cos\theta_c\cos\theta_2\\&\hspace{2cm}-\sin\theta_c\sin\theta_2\cos\phi_2\Big)+(1\leftrightarrow 2)\\
&g(1,2)\equiv\frac{1}{(1-\cos\theta_1\cos\theta_2-\sin\theta_1\sin\theta_2\cos\Delta\phi)}.
\end{aligned}\right.
\end{equation}
Since $\theta_c$ is small, we may Taylor expand $f(1,2,\theta_c)$ around $\theta_c=0$ to determine how the correlation function $\mathcal{I}(1,2)$ behaves for small $\theta_c$, 
\begin{equation}
\begin{aligned}
\mathcal{I}(1,2)&=f(1,2,\theta_c)\,g(1,2)-1, \\
&=\left (\sum_{n\in \mathbb{Z}} f_n(1,2)\,\theta_c^n\right )\,g(1,2)-1\\
&=\left (\frac{f_{-2}(1,2)}{\theta_c^2}+\frac{f_{-1}(1,2)}{\theta_c}+f_0(1,2)\right )g(1,2)\\&\hspace{0.7cm}-1+\mathcal{O}(\theta_c).
\end{aligned}
\end{equation}
In terms of the emitted gluons' angles, the series coefficients for the $f$ function are
\begin{equation}
\left \{\begin{aligned}
&f_{-2}(1,2)=4 (1- \cos \theta_1- \cos \theta_2+\cos \theta_1\cos \theta_2)\\
&f_{-1}(1,2)=2 \cos \phi_1 \sin \theta_1( \cos \theta_2-1)+(1\leftrightarrow 2)\\%&\hspace{1.6cm}+2 \cos \phi_2\sin \theta_2( \cos \theta_1 -1) \\
&f_0(1,2)=\frac{ 1+2 \cos \theta_1+2 \cos \theta_2-5 \cos \theta_1 \cos \theta_2}{3}.
\end{aligned}\right .
\label{coefficient_of_thetaC}
\end{equation}

Let us now expand these $f_i$ coefficients for small angles $\theta_i$. Up to $\mathcal{O}(\theta^3)$ the relations in \eq{coefficient_of_thetaC} become much simpler:
\begin{equation}
f_{-2}=f_{-1}=0\quad\textrm{and}\quad f_0=1- \cos \theta_1 \cos \theta_2.
\end{equation}
We may then write the two particle correlation function for the emission of two gluons collinear to the off-shell quark, after some simplification, as
\begin{multline}
\mathcal{I}(\theta_1,\theta_2,\Delta\phi)=\\ \frac{\sin\theta_1\sin\theta_2\cos\Delta\phi}{1-\cos\theta_1\cos\theta_2-\sin\theta_1\sin\theta_2\cos\Delta\phi}.
\end{multline}
We can find an even simpler expression by using the pseudorapidities $e^{-\eta_i}\equiv\tan(\theta_i/2)$ of the emitted gluons: 
\begin{equation}
\mathcal{I}(\Delta\eta,\Delta\phi)=\frac{\cos\Delta\phi}{\cosh\Delta\eta-\cos\Delta\phi}.
\label{correlation_in_rapidity}
\end{equation}
In equation \eq{correlation_in_rapidity} the correlation is expressed in a compact form, and the conformal invariance is explicitly manifest since it is expressed only in terms of angles. 

% \begin{figure}[!htbp]
%     \centering
%     \begin{subfigure}[b]{0.22\textwidth}

%        \resizebox{\textwidth}{!}{ \includegraphics{Images/twoparticle.pdf}}
%     \end{subfigure}
%     \begin{subfigure}[b]{0.25\textwidth}

%         \resizebox{\textwidth}{!}{\includegraphics{Images/CorrelationALICE.png}}
%     \end{subfigure}
%     \caption{The two particle correlations from (left) our non-Abelian, two gluon emission expression Eq.\ (\protect\ref{correlation_in_rapidity}) and from (right) central $pPb$ collisions as measured by the ALICE collaboration \protect\cite{Abelev:2012ola}. Note that the predicted two particle correlation plot on the left is not normalized, unlike the experimental result on the right.} 
%     \label{Correlation_plot}
% \end{figure}
% \begin{figure}[!htbp]
%     \centering
%     \includegraphics[width=0.8\columnwidth]{Images/twoparticle.pdf}
%     \includegraphics[width=\columnwidth]{Images/CorrelationALICE.png}
%     \caption{The two particle correlations from (top) our non-Abelian, two gluon emission expression Eq.\ (\protect\ref{correlation_in_rapidity}) and from (bottom) central $pPb$ collisions as measured by the ALICE collaboration \protect\cite{Abelev:2012ola}. Note that the predicted two particle correlation plot on the left is not normalized, unlike the experimental result on the right.} 
%     \label{Correlation_plot}
% \end{figure}
\begin{figure}[!htbp]
    \centering
    \includegraphics[width=\columnwidth]{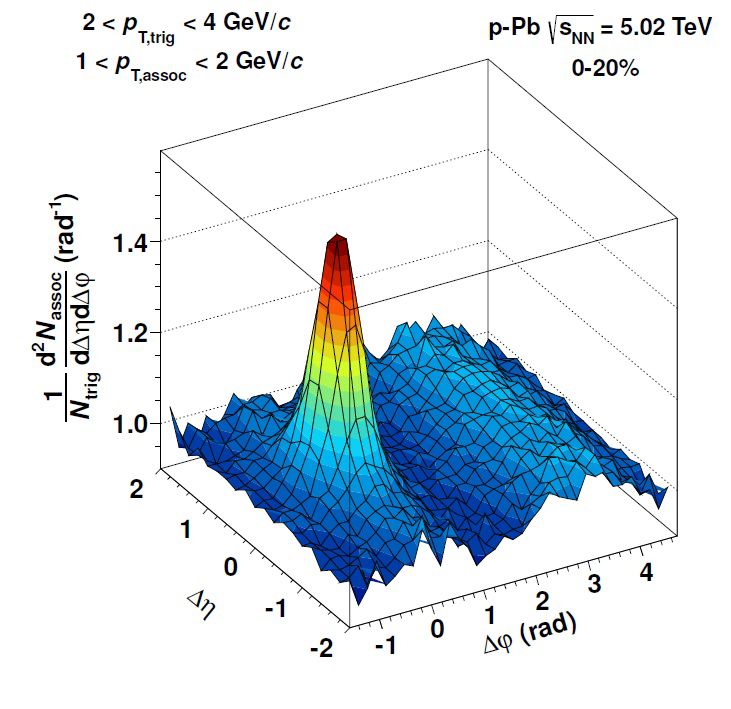}
    \includegraphics[width=0.8\columnwidth]{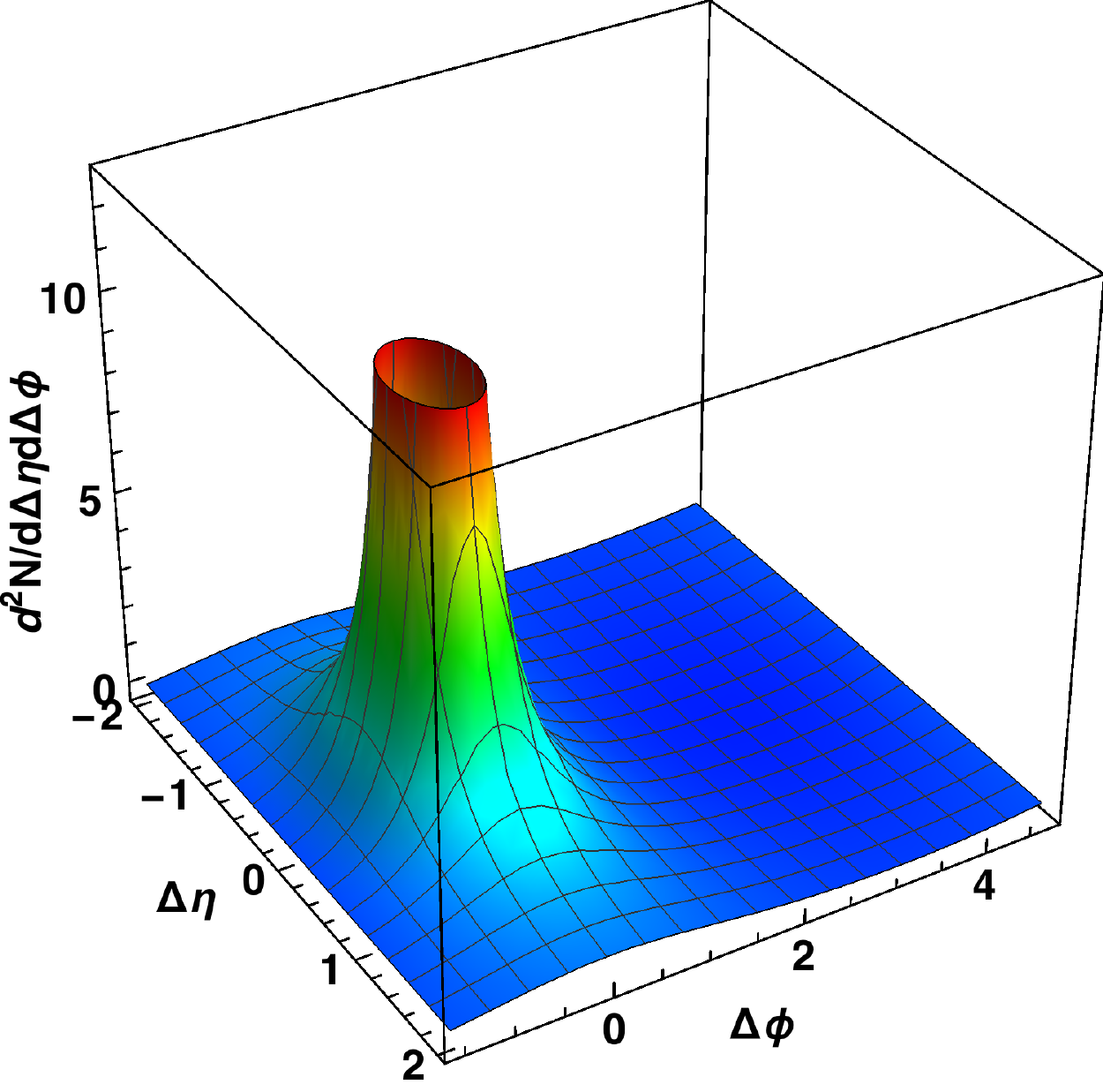}
    \caption{The two particle correlations from (top) central $pPb$ collisions as measured by the ALICE collaboration \protect\cite{Abelev:2012ola} and from (bottom) our non-Abelian, two gluon emission expression Eq.\ (\protect\ref{correlation_in_rapidity}). Note that the predicted two particle correlation plot on the bottom is not normalized, unlike the experimental result on the top.} 
    \label{Correlation_plot}
\end{figure}

\subsection{Two particle correlations phenomenology}
We would like to compare our two particle correlation result \eq{correlation_in_rapidity} with data.  We imagine a hadronic collision at a particle accelerator.  Then we can think of the incoming parton in \fig{angles} as inside one incoming hadron that is subsequently deflected by a small angle in the collision with the opposing hadron.  The incoming parton then radiates gluons at a small angle with respect to the incoming hadron.  A plot of the two particle correlations amongst the emitted quanta, \eq{correlation_in_rapidity}, are shown in the left hand plot of \fig{Correlation_plot}.  One can see that the near-side correlations are tight in angle $\phi$ but are long range in rapidity $\eta$.  We show in the right hand plot of \fig{Correlation_plot} the two particle correlation measurement from central $pPb$ collisions at ALICE \cite{Abelev:2012ola}, whose tight angular and long range in rapidity near-side correlation is very similar to our two particle correlation result.

Several comments are in order.  Our theoretical prediction is rigorously valid only for small angles of deflection and for small emission angles.  Nevertheless, small angles of emission still imply our result is valid for rapidities of $\eta\gtrsim0.7$.  Recent work also suggests that the result also holds for $2\rightarrow3$ scattering in which the incoming particles are scattered by a large $\sim\pi/2$ angle \cite{rabemananjara}.  While our correlation in rapidity is much larger than in angle, our result does have an exponential dropoff in $\Delta\eta$ which may be stronger than that observed in data.  

A quantitative comparison with data is difficult for several reasons.  First, our result formally diverges as $\Delta\eta$ and $\Delta\phi$ go to zero, whereas hadronization and detector effects mean that the ALICE data is finite for $\Delta\eta=\Delta\phi=0$.  Further, there is a lack of understanding of the specific mechanisms in small system collisions that lead to high multiplicity events.  The calculation presented here is based on the two gluon correlation from a single scattering.  However, in an actual high multiplicity $pp$ or $pA$ collision, it is currently unknown how many of measured particles originate from many semi-hard jets or from only a few very hard jets.  On the one hand, if the mechanism is dominated by semi-hard jets, our correlation will be suppressed.  On the other hand, the correlation presented in \fig{Correlation_plot} could be an underestimate if high multiplicity events are dominated by only a few very hard jets that yield significant numbers of correlated radiated quanta. 

\section{Induced \texorpdfstring{$v_2$}{v\_2}}
\label{sec:inducedv2}
The single inclusive distribution of radiated gluons is $\phi$ independent. 
However for the case of two gluon emission, the correlation between the two emitted gluons, \eq{correlation_in_rapidity}, depends on the azimuthal separation $\Delta\phi$, which will induce an azimuthal asymmetry.  

In heavy ion collisions, one associates a theoretical prediction of the azimuthal asymmetry with the experimentally accessible second Fourier coefficient of the angular distribution of measured particles, known as $v_2$, 
\begin{multline}
\left (\frac{dN}{d\eta dk d\phi}\right )\propto \\ 1+2v_1(\eta,k)\cos\phi+2v_2(\eta,k)\cos2\phi+\cdots.
\end{multline}
We would like to determine the non-flow contribution to the measured $v_2$ from the correlations between multiple gluon emissions, in this case for two gluons, $\mathcal{I}(\Delta\eta,\Delta\phi)$. In order to compute $v_2$ we need to perform a double Fourier decomposition in the angles of the two emitted gluons, $\phi_1$ and $\phi_2$, from which we can extract the average $v_2$ as was done in \cite{Kovchegov:2002cd},
\begin{equation}
\abraket{v_2}^2=2\pi\abraket{\cos2\phi_1\cos2\phi_2}_c.\label{v_2correlation}
\end{equation}

The quantity $\abraket{\cos2\phi_1\cos2\phi_2}_c$ on the right hand side of the above equation is the average of the product of the cosines with the non-Abelian correction to the Poisson independent emission approximated distribution, \eq{correcteddistribution},
\begin{equation}
\begin{aligned}
\abraket{v_2}^2=&\frac{2\pi K}{\abraket{N_{2}}}\int \mathcal{I}(\Delta\eta,\Delta\phi)\cos(2\phi_1)\cos(2\phi_2)d\Gamma,
\end{aligned}
\label{averagecorrelation}
\end{equation}
where $K$ is a constant factor that contains the coupling and color dependences of the two emissions. The above average is normalised by $\abraket{N_{2}}$, the average number of  gluon pairs emitted,
\begin{equation}
\abraket{N_{2}}=K\int \Big (1+ \mathcal{I}(\Delta\eta,\Delta\phi)\Big )d\Gamma,
\label{numberofpairs}
\end{equation}
with
\begin{equation}
d\Gamma=\!\!\!\prod_{i=\{1,2\}}\!\!\!d\phi_i d\eta_i\frac{dk_i}{k_i}.
\end{equation}

We want to evaluate the average $\langle v_2\rangle$ given in equation \eq{v_2correlation} for a pseudorapidity range $|\eta|<2.4$ and a transverse momentum range of $0.3<k<3$ GeV. The lower limit of the rapidity range is set by requiring small angle emission; the rapidity and momentum range was set to be as in \cite{Khachatryan:2014jra,Khachatryan:2016txc}.

Blind application of \eq{averagecorrelation} and \eq{numberofpairs} leads to an unsurprising IR divergence, since gluons are massless and indistinguishable.  This divergence will be eliminated by a careful examination of non-perturbative fragmentation physics.  In order to make progress without a full fragmentation function analysis, we introduce a simple approximation to the non-perturbative physics by requiring that our two gluons be emitted a sufficient distance apart in momentum space,
\begin{equation}
    R_{\eta\phi}=\sqrt{\Delta\eta^2+\Delta\phi^2}>R_{min},
\end{equation}
such that the gluons form separate hadrons.  We fix this minimum distance $R_{min}$ by the following estimate.  If we take the size of the pion to be in the range of $r_\pi\sim0.6$ fm \cite{Dally:1982zk,Beringer:1900zz}, the formation time of a pion in the lab frame as $\tau_\pi=\gamma/m_\pi$, then
\begin{equation}
  \label{rmin}
  R_{min} = 2\tan^{-1}(r_\pi/\tau_\pi) \sim 0.05-0.50
\end{equation}
for $p_T\sim1-4$ GeV$/c$ pions of mass 120 MeV. In addition to the minimal resolution $R_{min}$, we will constrain to $\left |\Delta\eta\right |>2$ in order to make a comparison with experimental measurements \cite{Khachatryan:2016txc}. (We will comment further below about the applicability of our calculation in this particular $\eta$ range.) We then have that

\begin{equation*}
\abraket{v_2}^2=\frac{2\pi K}{\abraket{N_{2}}}\int\mathcal{I}(\Delta\eta,\Delta\phi)\cos(2\phi_1)\cos(2\phi_2)\Theta(\eta,\phi)d\Gamma,
\end{equation*}
and
\begin{equation*}
\abraket{N_{2}}=K\!\!\int\Big (1+ \mathcal{I}(\Delta\eta,\Delta\phi)\Big )\Theta(\eta,\phi)d\Gamma,
\end{equation*}
with 
\begin{equation*}
\begin{aligned}
\Theta(\eta,\phi)&\equiv\Theta(\left |\Delta\eta\right |-2)\Theta(R_{\eta\phi}-R_{min}).
\end{aligned}
\end{equation*}
We then find the induced average azimuthal anisotropy is
\begin{equation}
\abraket{v_2}=0.13.
\label{v2}
\end{equation}
(The above $v_2$ is the same for the range different $R_{min}$ values in \eq{rmin} out to 5 decimal places.)

This value of $v_2$ is large compared to that measured in experiment, $v_2\sim0.05$ from experiments \cite{Khachatryan:2014jra,Khachatryan:2016txc}.  However, this value is the $v_2$ induced within a single jet of emitted particles from the two gluon emission from a single scattering event.  In a high multiplicity $pp$ or central $pPb$ collision, there can be multiple hard parton scatterings.  Gluons emitted from different scattered partons will be uncorrelated in angle; it is only the multiple gluons emitted from a single scattered parton that have angular correlations.  

In order to correctly predict the influence of the non-flow correlations presented here to the measured $v_2$, we must take into account the correct number of particles that are actually correlated.  The probability that two gluons randomly picked from all the radiated gluons belong to the same jet is
\begin{equation}
p_s=\frac{\#\textrm{Correlated pairs}}{\#\textrm{Possible pairs}}.
\end{equation}

If $N$ is the total number of emitted gluons for a given collision, then the number of possible particle pairs is given by 
\begin{equation}
\#\textrm{Possible pairs}=\left (\begin{matrix}
N\\2
\end{matrix}\right )
\end{equation}
while the number of correlated pairs is given by 
\begin{equation}
\#\textrm{Correlated pairs}=N_jN_c\times\left (\begin{matrix}
N_s\\2
\end{matrix}\right ),
\end{equation}
where $N_c$ is the number of hard scatterings per collision; $N_j$ the number of hard back-to-back jets per hard scattering, where we will take $N_j=2$; and $N_s$ is the average number of gluons radiated per hard jet.  For a central $pPb$ collision we can estimate $N_c\simeq14.8$ from an optical Glauber model \cite{Miller:2007ri} with $\sigma_{NN}=7.0$ fm$^2$ for $\surd s_{NN}=5.02$ TeV \cite{Adam:2015ptt}, and, with $\alpha_s=0.3$, we have
\begin{equation}
N_s=\frac{2\alpha_sC_F}{\pi} \int_0^{2\pi} \frac{d\phi}{2\pi} \int_{0.7}^4d\eta\int_{0.3}^3\frac{dk}{k}\approx 2.
\end{equation}
If we assume parton-hadron duality \cite{Melnitchouk:2005zr}, in which the number of produced hadrons is equal to the number of produced partons, then we find that the number of measured charged hadrons $N$ is 
\begin{equation}
N=\frac{2}{3}N_jN_c (1+N_s)\approx 60.
\end{equation}

The factor of $2/3$ comes from assuming the produced hadrons are dominated by pions, of which only 2/3 are charged. Therefore what we will call the natural reduced average $v_2$ for central $pPb$ collisions is 
\begin{equation}
\overline{\abraket{v_2}}\equiv p_s\abraket{v_2}=0.01.
\end{equation}

What we would like to compare to is $v_2(N)$, where $N$ can range up to values $\sim250$.  It is currently unclear what mechanism(s) are responsible for such high multiplicity events for the $pp$ and $pPb$ small collision systems.  We can make two different simple assumptions for the origin of these high multiplicity collisions.  First, that the high multiplicity is the result of a few extremely high energy jets that yield a very large number of semi-hard particles through a very large number of branchings; i.e.\ we can fix the $N_c$ but allow $N_s$ to vary to yield the measured $N$.  Second, we can fix the number of emitted quanta $N_s$ but allow the number of hard collisions $N_c$ to vary in order to yield the measured $N$.  

In the case where number of tracked hadrons $N$ is dominated by decay products of branchings from very high energy original partons, we can fix the number of collision $N_c$ and let the number of radiated gluons scale with number of measured hadrons, from which
\begin{equation}
N_s=\frac{N}{(2/3)N_j N_c}-1.
\end{equation}

In the case where $N$ is dominated by decay products of very many semi-hard  collisions, we keep $N_s$ fixed and let the number of hard collision scales with $N_c$, from which
\begin{equation}
N_c=\frac{N}{(2/3) N_j (1+N_s)}.
\end{equation}

\begin{figure}[h!]
\centering
\resizebox{0.5\textwidth}{!}{
\includegraphics{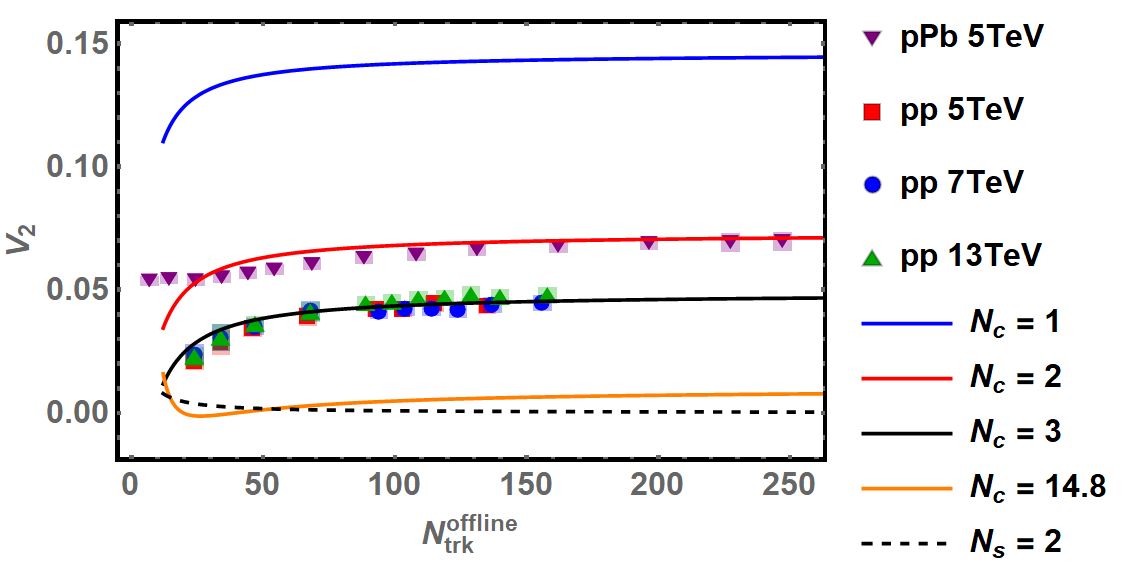} }
\caption{Comparison of \redvtwo{} to the $v_2$ measured by CMS for $pp$ and $pPb$ with  $|\eta|<2.4$, $|\Delta\eta|>2$,  and $0.3<p_T<3$ GeV as in \cite{Khachatryan:2014jra,Khachatryan:2016txc}.}
\label{Fig_v2}
\end{figure}

Using these two simple approximations, we compare in \fig{Fig_v2} the scaling of $v_2(N)$ from our two particle correlation prediction and data from the CMS experiment in $pp$ and $pPb$ collisions \cite{Khachatryan:2014jra,Khachatryan:2016txc}.  The solid curves correspond to the first assumption, in which the number of high energy jets is fixed with the number of particles per jet allowed to vary in order to produce the necessary $N$.  We show the results for $N_c=1,\,2,\,3,$ and $14.8$.  Recall that $N_c=14.8$ is the expected number of binary collisions from the Optical Glauber model.  The dashed curve corresponds to the second approximation, in which $N_s=2$ is fixed and $N_c$ varies in order to match the measured $N$.   

Not surprisingly, one sees that for the first model, in which $N_c$ is fixed, \redvtwoN{} decreases as one increases $N_c$: as $N_c$ increases there are more uncorrelated jets from which to pick out two particles, decreasing the anisotropy.  Also unsurprisingly, as $N$ increases for fixed $N_c$, \redvtwoN{} increases: more and more particles are necessarily correlated with the fixed number of jets in order to add up to the given $N$ for a given fixed $N_c$.  Similarly, for the fixed $N_s=2$, \redvtwoN{} decreases with increasing $N$ as the number of jets required to reach $N$ for the fixed $N_s$ increases.

Compared to data, there is incredibly good agreement between the predicted \redvtwo{} and the $pp$ $v_2(N)$ from CMS \cite{Khachatryan:2014jra} for the fixed number of hard scatterings $N_c=3$.  The agreement is especially intriguing since the number of hard collisions required is reasonable and also because the multiplicity dependence is qualitatively so similar.  Of course, the number of hard scatterings in any one $pp$ collision fluctuates, but it is very easy to arrive at a \redvtwoN{} averaged over a reasonable distribution of hard scatterings that is essentially the same as the $N_c=3$ result shown.

Let us now compare our \redvtwoN{} to the $pPb$ data from CMS \cite{Khachatryan:2016txc}.  One can see again an excellent agreement between the fixed $N_c$ model and data; but, because the agreement is for a smaller $N_c=2$ than for $pp$ collisions, and because the expected typical number of hard scatterings is more like $N_c=14.8$, the agreement must be spurious.  However, since the expected number of measured charged particles from the Optical Glauber model is $\sim60$ and there are many $pPb$ collisions with far fewer produced particles, we see again a lack of understanding of the multiplicity distribution: it is not clear that the fixed $N_c=14.8$ result is a reasonable comparison to $pPb$ $v_2(N)$ data.  Perhaps it would make more sense to compare with a smaller fixed $N_c$, in which case a not insignificant fraction of the $v_2(N)$ measured by CMS is actually due to non-flow effects from the non-Abelian correlation of particle production in QCD.

A comment is in order regarding the applicability of our correlations calculation and the data presented by CMS.  Our result is strictly valid only for small angle emissions, which require that $|\eta|>0.7$.  Since CMS requires that the two particles used in their correlations measurement satisfy $|\eta|<2.4$ \emph{and} $\Delta\eta>2$, at least one of the particles cannot come from small angle emission.  As such our comparison to data is to some extent an uncontrolled approximation, as it is unclear how large the corrections would be if we were to include contributions from wide angle scattering.  However, if we restrict our calculation to a region in which it is fully under control, for example for $0.7<\eta<4$ with $\Delta\eta>2$ we find $v_2=0.17$, which is within 30\% of the value quoted before in \eq{v2}, $v_2=0.13$; i.e.\ the contribution to $v_2$ from the non-Abelian correlation computed here appears relatively insensitive to the exact details of the chosen $\eta$ range, thus boosting our confidence in the application of our calculation outside of its strict range of applicability.  Further,  our result depends only on $\Delta\eta$, so it is likely that our result only depends on the difference $\eta_{max}-\eta_{min}$ rather than the individual values of $\eta$.

\section{Conclusions}
\label{sec:conclusions}
In this paper we studied the emission of one and two gluons from an off-shell quark.  We introduced the spinor helicity formalism and the maximal helicity violating (MHV) technique, which provide a powerful alternative to the usual Feynman diagram tools for computing multiple gluon emission amplitudes.  We reviewed the trivial derivation of the emission of a single gluon using MHV.  We then computed the cross section for the emission of two gluons.  We found that for non-Abelian QCD, unlike in QED, the emission probability for two gluons is not simply an independent Poisson convolution of the single inclusive gluon emission probability.  We explicitly derived the non-Abelian correlations between the two emitted gluons, \eq{GluonCorrelation}, which simplifies to the manifestly conformal \eq{correlation_in_rapidity} in terms of only the difference in angle, $\Delta\phi$, and rapidity, $\Delta\eta$, between the two emitted gluons for the case of collinear radiation.  

We then investigated the phenomenological relevance of our results.  A direct comparison between the two gluon correlations we computed and recent ALICE data from central $pPb$ collisions \cite{Abelev:2012ola} shows a surprisingly good qualitative agreement: both distributions display a tight correlation in angle and broad correlation in rapidity.  The conformality of our correlations prediction, \eq{correlation_in_rapidity} implies that the shape of the two particle correlations measured in collisions without medium modification will be independent of the momentum cuts made.

We also studied the influence of the non-Abelian correlations amongst multiple gluon emission on the measured $v_2$ in high multiplicity $pp$ and $pPb$ collisions.  We found a remarkable agreement between our prediction and the measured semi-hard particle $v_2$ as a function of multiplicity in $pp$ collisions, without having to resort to the perhaps exotic application of hydrodynamics to such a small system.  The non-Abelian, non-flow contribution to $v_2$ decreases as the number of jets increases, but a sizable portion of the $v_2$ measured in $pPb$ collisions may come from non-Abelian semi-hard jet correlations.   

The origin of the correlations from multiple gluon emission is very different from those of \cite{Dumitru:2010iy,Ozonder:2016xqn,Dusling:2017dqg}, which also describe a contribution to two particle correlations measured in high multiplicity collisions of small systems.  In \cite{Dumitru:2010iy,Ozonder:2016xqn,Dusling:2017dqg}, the correlations are due to multiple interactions between the incoming particles; in the language of those papers, the correlations are due to multiple interactions between the projectile and target.  Our multiple gluon emission correlations, however, are insensitive to the details of the process that produces the original off-shellness, which could be generated by a single scattering interaction.

Our two gluon correlation calculation is useful beyond the provocative comparison to the measured two particle correlations and $v_2$ in small system collisions with high multiplicity.  Our correlations calculation can also provide a new constraint on Monte Carlo generators, which are of critical use in computing the QCD background in new physics searches.  And we anticipate that the MHV technique presented here can be fruitfully applied in other areas of heavy ion phenomenology, most obviously in computing the non-Abelian, non-Poisson multi-gluon corrections to radiative energy loss calculations.

\begin{acknowledgements}

The authors wish to thank the SA-CERN Collaboration and the National Research Foundation for generous support.  The authors wish to thank Miklos Gyulassy and Jan Fiete Grosse-Oetringhaus for useful discussions.

\end{acknowledgements}

% For tables use

% Or use

%
% BibTeX users please use
% \bibliographystyle{}
% \bibliography{}
%
% Non-BibTeX users please use

\end{document}